\documentclass[12pt,preprint]{aastex}

\def\lesssim{\mathrel{\hbox{\rlap{\hbox{\lower4pt\hbox{$\sim$}}}\hbox{$<$}}}}
\def\gtrsim{\mathrel{\hbox{\rlap{\hbox{\lower4pt\hbox{$\sim$}}}\hbox{$>$}}}}

\slugcomment{}
\shorttitle{Lightman-Eardley Instability}
\shortauthors{Hirose et al.}

\begin{document}

\title{Turbulent Stresses in Local Simulations of
Radiation-Dominated Accretion Disks, and the
Possibility of the Lightman-Eardley Instability}

\author{Shigenobu Hirose}
\affil{Institute for Research on Earth Evolution, JAMSTEC, Yokohama,
Kanagawa 236-0001, Japan}

\author{Omer Blaes}
\affil{Department of Physics, University of California, Santa Barbara,
   Santa Barbara CA 93106}

\and

\author{Julian H. Krolik}
\affil{Department of Physics and Astronomy, Johns Hopkins University,
    Baltimore, MD 21218}

\begin{abstract}
We present the results of a series of radiation-MHD simulations of a local
patch of an accretion disk, with fixed vertical gravity profile but with
different surface mass densities and a broad range of radiation to gas
pressure ratios.  Each simulation achieves a thermal equilibrium that lasts
for many cooling times.  After averaging over times long compared to a
cooling time, we find that the vertically integrated stress is approximately
proportional to the vertically-averaged total thermal (gas plus radiation)
pressure.  We map out---for the first time on the basis of explicit
physics---the thermal equilibrium relation between stress and surface
density: the stress decreases (increases) with
increasing surface mass density when the simulation is radiation (gas)
pressure dominated.  The dependence of
stress on surface mass density in the radiation pressure dominated regime
suggests the possibility of a Lightman-Eardley inflow instability, but global
simulations or shearing box simulations with much wider radial boxes will
be necessary to confirm this and determine its nonlinear behavior.

\end{abstract}


\section{Introduction}

It has long been known that models of optically thick, geometrically thin
accretion disks based on the alpha stress prescription of \citet{sha73}
are subject to thermal and inflow (``viscous'') instabilities when
the vertically-averaged radiation to gas pressure ratio exceeds 3/2
\citep{le74,shi75,sha76}.  Such radiation pressure dominated accretion disks are
expected to be relevant for luminous active galactic nuclei and quasars as well
as for thermal states of X-ray binaries.  However, with one possible exception,
there has never been clear observational evidence, or even observational
motivation, for the existence of these instabilities in these sources.
This is in marked contrast to the
cases of dwarf novae and soft X-ray transients, where thermal instabilities
in the disk associated with hydrogen ionization, not radiation pressure,
are central to explaining the observed outbursts \citep{las01}.  The X-ray
binary GRS~1915+105 does exhibit recurrent outburst behavior that has been
modelled as being due to radiation pressure driven instabilities \citep{bel97},
but it is far from clear that this is the correct explanation.  This
source is the brightest among Galactic black hole X-ray binaries, and
spends considerable time at super-Eddington luminosities \citep{don04}.
Other black hole X-ray binaries commonly reach high enough Eddington ratios
for instabilities to exist according to standard accretion disk
theory, but do not exhibit similar variability.

It is now widely suspected that accretion stresses in black hole accretion
disks are due to turbulence related to the nonlinear growth of the
magnetorotational instability (MRI) \citep{bal98}.  It
is computationally feasible to perform thermodynamically consistent,
radiation MHD simulations of this turbulence in local
patches of accretion
disks.  These stratified shearing box simulations fully capture grid-scale
numerical losses of energy as heat and also account for radiative heat losses
within the flux-limited diffusion approximation \citep{hir06}.  Such
simulations have now been performed for a broad range of radiation to gas
pressure ratios, and in each case an approximate thermal equilibrium has been
established lasting for many cooling times \citep{hir06,kro07,hir09}.
No sign of the radiation pressure thermal
instability is present, even at radiation to gas pressure ratios well above
the instability threshold of the standard alpha model (\citealt{hir09}; see
also \citealt{tur04}).

Thermal stability exists because the stress-pressure relation assumed by
the standard alpha-model is only established on time scales longer than
the thermal time.\footnote{In a prescient comment, \citet{le74}
suggested that the alpha prescription might only be valid on slow time
scales, which they identified as being of order the inflow time and longer.}
The causal direction of the relation is from stress
to pressure, not from pressure to stress.  Turbulence is chaotic and
results in a highly fluctuating dissipation rate.  It is that dissipation
that ultimately changes the pressure, but that pressure response is
only established after a thermal time.  An upward fluctuation in pressure
does not result in an upward stress response on this time scale as the
causal direction is the other way round.  Hence there is no positive
feedback loop on the thermal time scale that would result in a thermal
runaway \citep{hir09}.

This still leaves open the question of the slower inflow instability.
Mass and angular momentum conservation imply that radial mass transport
in an accretion disk with a local turbulent stress is governed by the
equation \citep{le74,lyn74}
\begin{equation}
{\partial\Sigma\over\partial t}={1\over r}{\partial\over\partial r}\left[
{1\over\ell^\prime}{\partial\over\partial r}(r^2W_{r\phi})\right],
\label{eqdsigmadt}
\end{equation}
where $\Sigma$ is the local surface mass density, $r$ is the radius,
$\ell^\prime\equiv d\ell/dr$ is the radial derivative of specific angular
momentum, and $W_{r\phi}$ is the vertically integrated turbulent stress.
For geometrically thin disks,
the inflow time is much longer than the thermal time, so both thermal
and vertical hydrostatic equilibrium should be maintained over the time
scales associated with mass transport.
Assuming the disk is optically thick and cools through radiative losses
(which is the case for all of our stratified shearing box simulations
so far; \citealt{hir06,kro07,hir09}), then radiative, hydrostatic,
and thermal equilibrium imply that the vertically-averaged stress in a
radiation pressure dominated disk is simply given by
\begin{equation}
\tau_{r\phi}={c\Omega^2\over \kappa r|\Omega^\prime|},
\label{eqtaurphi}
\end{equation}
where $\Omega$ is the angular velocity in the disk, $c$ is the speed of
light, and $\Omega^\prime\equiv d\Omega/dr$ is the shear \citep{sha76}.
The opacity $\kappa$
is generally dominated by electron scattering in this regime, and is
therefore constant.  Hydrostatic equilibrium implies that the disk
half-thickness is $H\sim2P/(\Omega^2\Sigma)$, where $P$ is the midplane
pressure.  Hence at
a particular radiation pressure dominated radius,
\begin{equation}
W_{r\phi}\sim2H\tau_{r\phi}\propto{P\over\Sigma}.
\end{equation}
A standard alpha disk model with $\tau_{r\phi}=\alpha P$ implies from
equation (\ref{eqtaurphi}) that $P$ is independent of $\Sigma$, so
$\partial(W_{r\phi})/\partial\Sigma<0$. Equation (\ref{eqdsigmadt})
therefore represents a diffusion equation with a {\it negative} diffusion
coefficient. Unstable growth of surface density enhancements and rarefactions
would therefore result \citep{le74,lig74a,lig74b}.

Like the fictitious thermal instability, however, the reality of the inflow
instability has always been questionable.
For example, as pointed out by \citet{le74} themselves, a
stress proportional to the gas pressure alone would produce an inflow stable
disk.  On the other hand, the stratified shearing box simulations appear to be
consistent with total thermal pressure scaling with stress on supra-thermal time
scales \citep{hir09}, a fact that we will demonstrate much more explicitly
in this paper.

A plot of vertically integrated stress $W_{r\phi}$ versus surface density for
a range of thermal equilibria at a fixed radius within a disk would suggest
inflow stability or instability depending on whether the slope is positive
or negative, respectively.  Because thermal equilibrium implies that the
local radiation flux $F$ emerging from each face of the disk is given
by $F=W_{r\phi}r|\Omega^\prime|/2$, a plot of $F$ (or effective temperature)
versus surface density may be used in the same way.  Such ``S-curve" plots
are commonly used to investigate the hydrogen ionization-driven disk
instabilities in dwarf novae and soft X-ray transients
(e.g. \citealt{sma84,las01}).

Using radiation MHD simulations of stratified shearing boxes, we have now
mapped out the stress versus surface density thermal equilibrium curve for
a wide range of radiation to gas pressure ratios at a fixed radius in a disk
around a stellar mass black hole.  This is the first time that this curve
has been drawn on the basis of explicit physical mechanisms, rather than
phenomenological estimates.  While there are large fluctuations
which produce an inherent scatter in the thermal equilibrium curve, the
results are consistent with the standard alpha disk model.  The inflow
instability might therefore be present even in MRI turbulent disks.

This paper is organized as follows.  In section 2 we provide a brief overview
of the numerical parameters and properties of the simulations.  In section 3
we discuss the stress-pressure relation and demonstrate that average total
thermal pressure, rather than a pressure which singles out gas pressure as
being special in some way, is best correlated with average stress.
In section 4, we summarize the resulting thermal equilibria on a
stress-surface density diagram, and discuss the possible implications.
We summarize our conclusions in section 5.

\section{Simulations}

The radiation MHD equations for stratified shearing boxes, and the numerical
methods we use to solve them, have been described in detail by \citet{hir06}
and \citet{hir09}, and references therein.  Grid scale losses of mechanical
and magnetic energy are
fully captured as heat in the gas, and gas and radiation exchange heat through
Planck mean free-free absorption and emission and Compton scattering.  Radiation
transport is treated through flux-limited diffusion.

All the simulations were run with an angular velocity $\Omega=190$~s$^{-1}$,
corresponding to a radius of $30GM/c^2$ around a 6.62~M$_\odot$ Schwarzschild
black hole.  Different total surface mass densities were chosen for
each simulation in order to map out the stress-surface density relation.
Table \ref{simparams} summarizes the parameters of each
simulation.  The $x$, $y$, and $z$ axes correspond to the radial, azimuthal,
and vertical directions, respectively.  The simulations were initialized
in hydrostatic and thermal equilibrium under an assumed initial vertical
profile of dissipation per unit volume proportional to the density divided
by the square root of the optical depth measured from the nearest surface.
Once the simulation starts, the dissipation is thereafter self-consistently
determined from the turbulent dynamics.  Each simulation started with a
weak magnetic field consisting of a twisted azimuthal flux tube located
at the center of the box.  We refer the reader to \citet{hir09} for more
details.

The vertical box height $L_z$ of the simulations was chosen so that two
conditions would be satisfied \citep{hir09}.
First, the total surface mass density changes by less than a few percent
due to vertical mass loss and mass creation by the density floor of the
simulation.  Second, the MRI is always well-resolved in the midplane regions.
These conditions were checked {\it a posteriori}.  The upper and lower
photospheres are always within the simulation domain, although the optically
thin regions do not necessarily have a tremendous vertical extent.  This
should not significantly affect the emergent flux, which is what really
matters.  Because the radiation energy density in the optically thin regions
is nearly independent of height, the radiation energy density at the
photospheres should depend only weakly on box height.  It is this radiation
energy density that provides the effective boundary condition for the
radiation diffusion equation in the optically thick regions where the vast
majority of the dissipation occurs, so the emergent flux should be
well-determined.  Nevertheless, we warn the
reader that we have not demonstrated numerical convergence with respect
to variations in the simulation box dimensions.

Six of the simulations (0211b, 0519b,
1112a, 1126b, 0520a, and 0320a) were initialized as radiation pressure
dominated, while two (090304a and 090423a) started out gas pressure dominated.
(Due to an initialization error, simulations 090304a and 090423a actually
had an angular velocity $1.6\%$ smaller than the others:
187~rad~s$^{-1}$ rather than 190~rad~s$^{-1}$.  We do not believe this
significantly affects our conclusions, as the fluctuations in stress
and radiative cooling are considerably larger than this.)
The results of simulations 1112a and 1126b were discussed extensively in
\citet{hir09}.

All of the simulations share many of the properties that we discussed
in detail in previous papers on radiation MHD simulations of stratified
shearing boxes \citep{hir06,kro07,bla07,hir09}.  The subphotospheric regions
consist of a magnetorotational turbulent zone in the midplane regions which
is supported against gravity by gas and radiation pressure gradients.  Further
out, magnetic forces dominate or contribute substantially, and Parker
instability dynamics, rather than MRI turbulence, appear to control the
structure of the outer layers.

\begin{figure}
\plotone{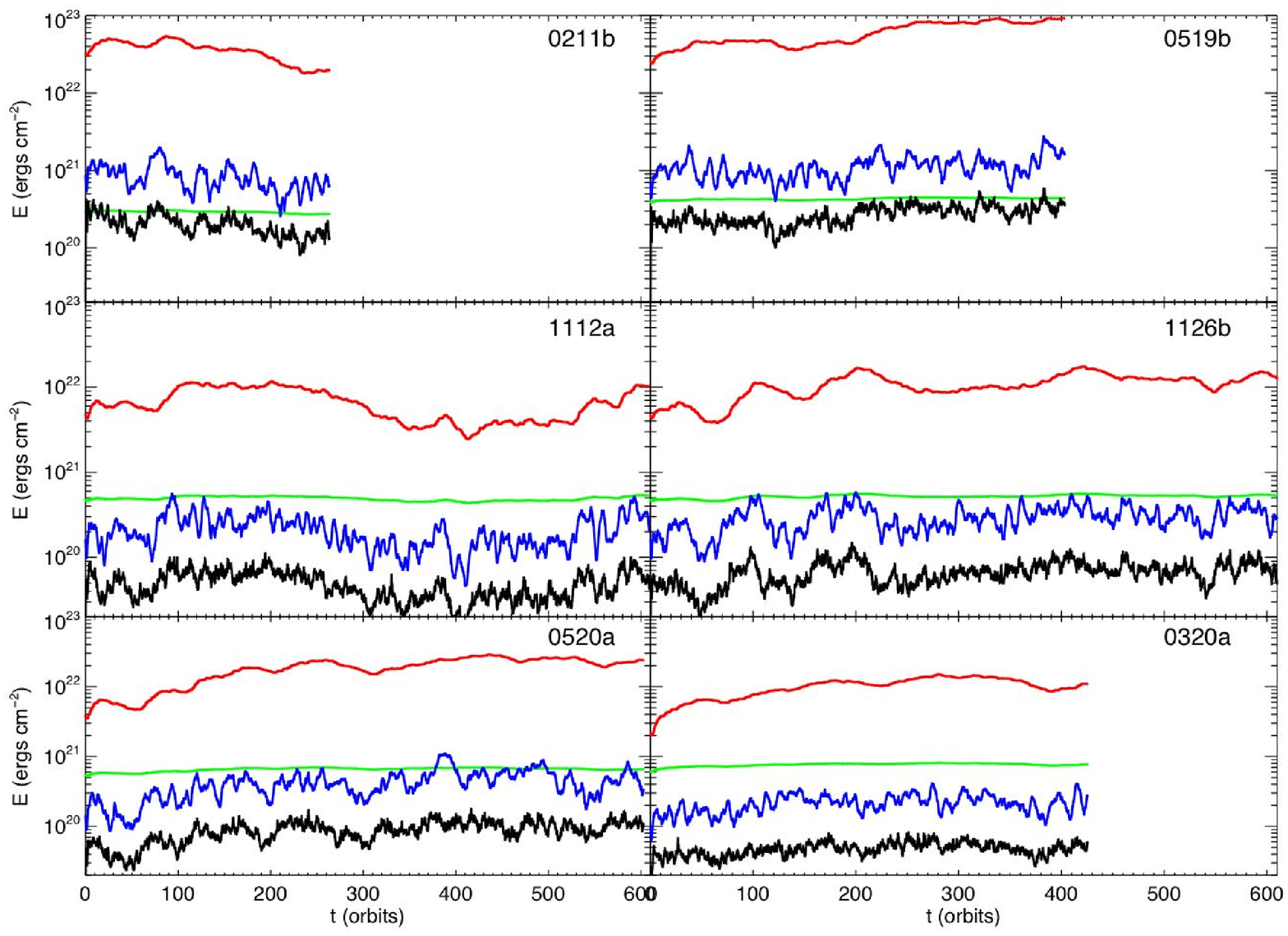}
\caption{Time history of the box integrated radiation internal energy (red),
gas internal energy (green), magnetic energy (blue) and turbulent kinetic
energy (black) for each of the radiation pressure dominated simulations.
\label{fig:thermalhistory}}
\end{figure}

\begin{figure}
\plotone{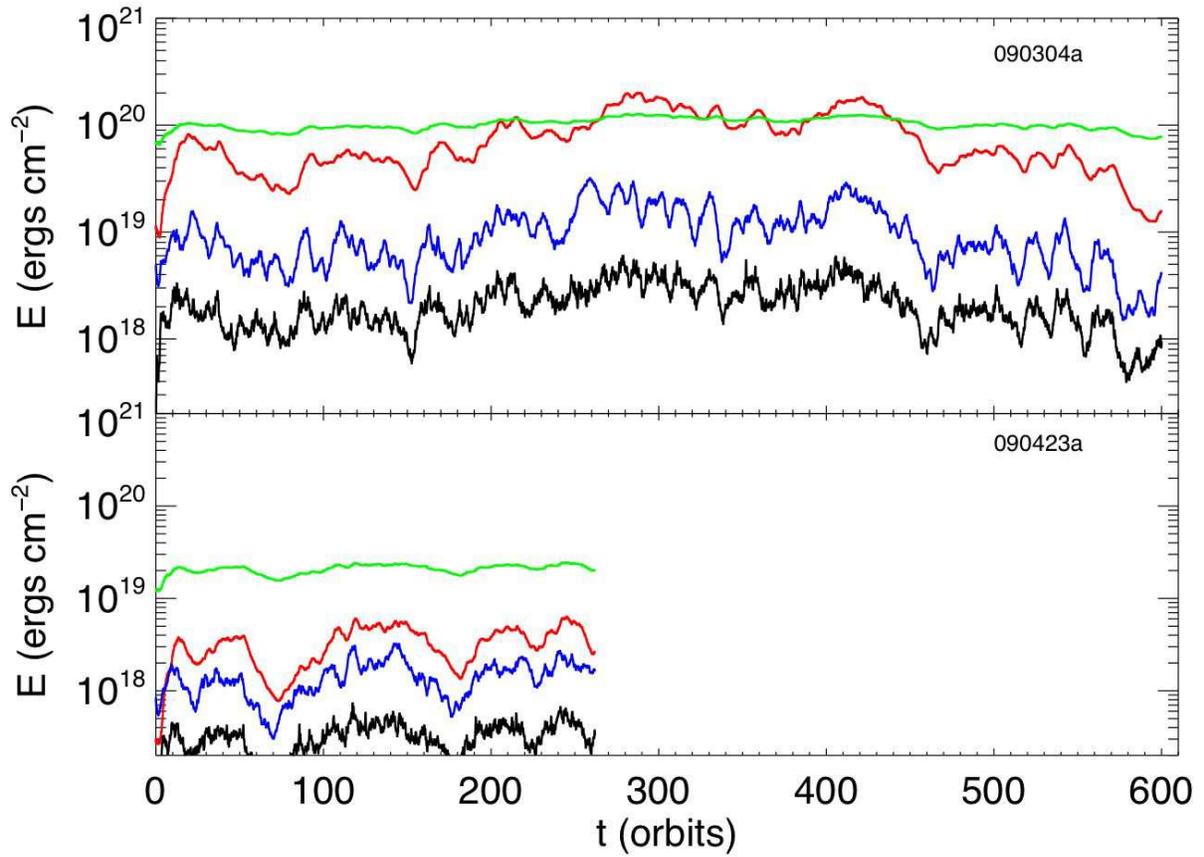}
\caption{Same as Figure \ref{fig:thermalhistory} except for simulations
090304a and 090423a, which are not radiation pressure dominated.
Note the much smaller energies on the vertical axis.
\label{fig:thermalhistorygasdom}}
\end{figure}

Figure \ref{fig:thermalhistory} shows the thermal and turbulent energy content
in the box as
a function of time for each of the radiation pressure dominated simulations.
The thermal history of the more gas-dominated simulations are shown
in Figure \ref{fig:thermalhistorygasdom}.  Defining the instantaneous
thermal time as the total radiation and gas internal energy divided by the
emergent radiative flux on both vertical faces of the box, the thermal time
averaged over the duration of each of the simulations
ranges from a minimum of 6 orbits for 090423a to a maximum of 24 orbits for
0519b.  All the simulations have reached
an approximate thermal equilibrium, albeit with long time scale fluctuations.
There is no evidence for the thermal instability predicted by classic alpha
disk models \citep{sha76}, in spite of the fact that the time-averaged
ratio of vertically-averaged radiation pressure to vertically-averaged gas
pressure is as high as 70 in the case of simulation 0519b.

\section{The Stress-Pressure Relation}

A number of authors have suggested alternative
stress prescriptions in which the accretion stress is proportional to gas
pressure or some
combination of gas and radiation pressures, rather than total thermal
pressure (gas plus radiation), in part to stabilize the radiation pressure
dominated portion of
black hole accretion disks \citep{sak81,ste84,bur85,szu90,mer02,mer03}.
\citet{sak89} have even argued that standard
alpha disk models are inconsistent in the radiation pressure dominated
regime, as magnetic fields that are strong enough to provide accretion
stress would be too buoyant to be retained by the disk.  Their argument has
two flaws, however.  First, they assumed that the magnetic field consisted
of discrete flux tubes, rather than being more continuously distributed
throughout the plasma.  Second, they were unaware at the time of magnetic
field generation by the MRI.  Note that there is no indication in the energy
histories shown in Figure \ref{fig:thermalhistory} that the magnetic and
turbulent kinetic energies are limited by the gas internal energy.  Indeed,
in the two most radiation pressure dominated simulations (0211b and 0519b),
the magnetic energy almost always exceeds the gas internal energy,
and even the turbulent kinetic energy can occasionally be larger than the
gas internal energy.

Based on the behavior of the stress to thermal pressure ratio within simulation
1112a, we argued in the past that a prescription where stress and total
thermal pressure are proportional to each other (on time scales longer than
the thermal time) is a superior description
of the simulation behavior than
alternatives where the pressure is taken to be just the gas pressure alone or
the geometric mean of radiation and gas pressure
(see Fig. 10 of \citealt{hir09}).  A similar conclusion can be drawn when
data is compared across the simulations we are considering here.

For each of our simulations, we computed the vertically-averaged stress
$\tau_{r\phi,{\rm av}}$ and the box-average of various measures of pressure
$P_{\rm av}$:  the sum of the radiation pressure $P_{\rm rad}$ and gas
pressure $P_{\rm gas}$, the geometric mean $(P_{\rm rad}P_{\rm gas})^{1/2}$
of these pressures, and just the gas pressure.  We then computed the time
average of the ratio of these spatial averages of stress and pressure,
$\alpha\equiv<\tau_{r\phi,{\rm av}}/P_{\rm av}>$, and compared them to the
time average of the ratio of spatially averaged radiation and gas pressure,
$<P_{\rm rad,av}/P_{\rm gas,av}>$.  These time averages were computed
over the entire duration of each simulation, excluding the first 10 orbits
as the MRI was still in its growth phase.

\begin{figure}
\plotone{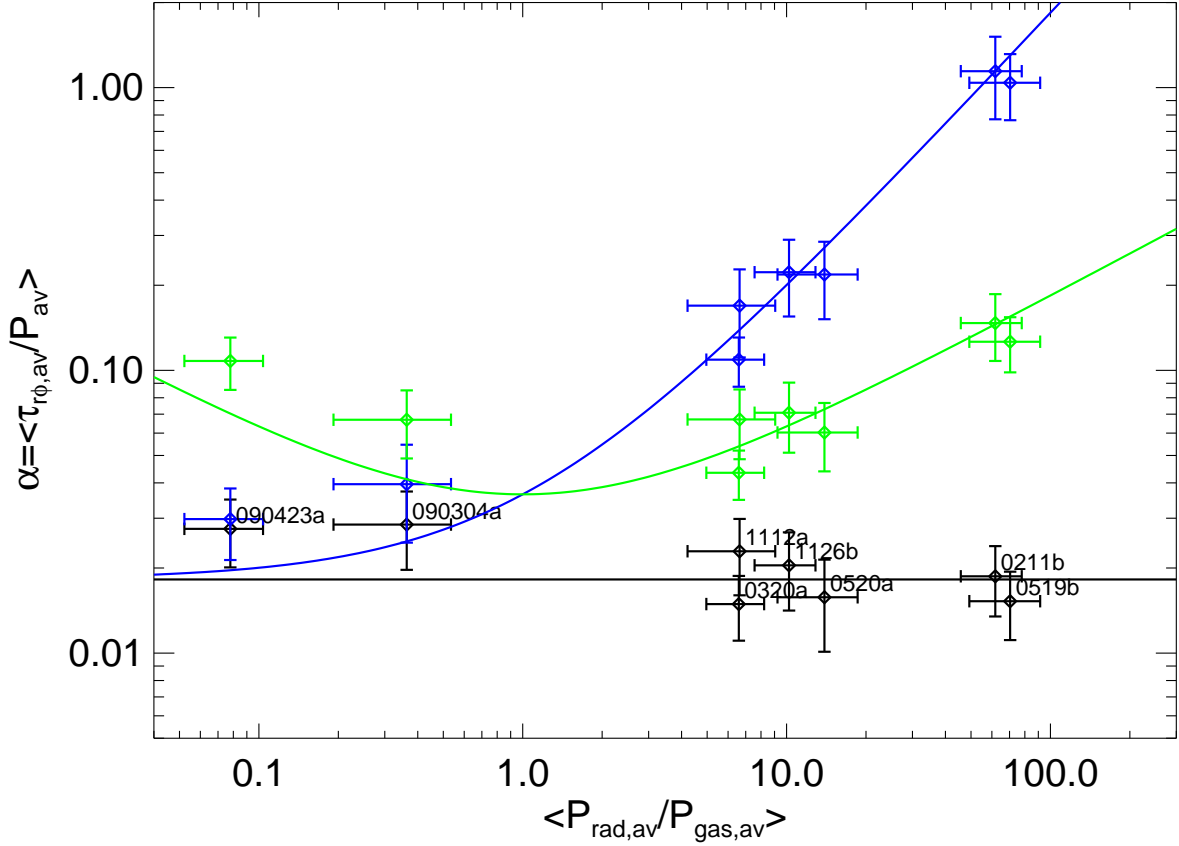}
\caption{Measured values of the stress parameter $\alpha$ as a function of
the time-averaged ratio of the box-averaged radiation pressure to the
box-averaged gas pressure.  The black points define $\alpha$ as the
time-averaged ratio of the vertically averaged stress to the box-averaged
total thermal pressure.  The blue and green points define $\alpha$
in the same way except with total thermal pressure replaced by gas pressure
and the geometric mean of gas and radiation pressure, respectively.  Both
horizontal and vertical error bars indicate one standard deviation in the
time averages.  The horizontal black line indicates the weighted mean of
$\alpha$ for the total pressure stress prescription.  Assuming that stress
really scales with total thermal pressure, the green and blue curves show
how $\alpha$ would then behave under the other stress prescriptions.
\label{fig:alphavspradopgas}}
\end{figure}

The results are shown in Figure~\ref{fig:alphavspradopgas}.  The black points 
show the results for $P_{\rm av}$ defined as the total pressure, and are
clearly closest to being independent of the radiation to gas pressure ratio.
(A linear fit to the dependence of these values of $\alpha$ on the radiation
to gas pressure ratio gives a slightly negative slope which is consistent with
zero within one standard deviation of the slope determination.)
A weighted average of these values gives $\bar{\alpha}=0.018\pm0.002$. The green
curve shows what one would obtain if stress and total pressure were proportional
with this ratio, but we redefined $\alpha$ in terms of the geometric mean of
radiation and gas pressure, i.e.
\begin{equation}
\left<{\tau_{r\phi,{\rm av}}\over P_{\rm av}}\right>\simeq\bar{\alpha}
\left[\left<{P_{\rm rad,av}\over P_{\rm gas,av}}\right>^{1/2}+
\left<{P_{\rm rad,av}\over P_{\rm gas,av}}\right>^{-1/2}\right].
\end{equation}
The blue curve shows the same thing for the gas pressure stress prescription,
\begin{equation}
\left<{\tau_{r\phi,{\rm av}}\over P_{\rm av}}\right>\simeq\bar{\alpha}\left(
1+\left<{P_{\rm rad,av}\over P_{\rm gas,av}}\right>\right).
\end{equation}
These curves clearly explain the trends seen in the data for these alternative
stress prescriptions.

Our results are therefore most consistent with the standard alpha prescription
involving total thermal pressure.  Recent axisymmetric global radiation MHD
simulations by \citet{ohs09} also reach a similar conclusion.  We emphasize,
however, that such a prescription should still be treated with caution.
There is no obvious reason that $\alpha$ should be a universal constant, and
we do not know what physics is determining its value in our simulations.  It
is possible, for example, that $\alpha$ is a function of $\Omega$ or $M$, as
we have not varied these parameters at all; global simulations \cite{hk01,hk02}
have shown that it can be a function of radius when the underlying orbital
dynamics change.
In fact, it is rather surprising that it is as constant as it is when defined
in terms of total thermal pressure.  Moreover,
as we emphasized in \citet{hir09}, the fact that stress is in any way related
to pressure is really because turbulent dissipation heats the plasma.  Only
when averaged over height and averaged over many thermal times does the
standard alpha prescription provide an adequate description of the fact that
pressure is correlated with stress.

\section{The Stress-Surface Density Relation}

\begin{figure}
\plotone{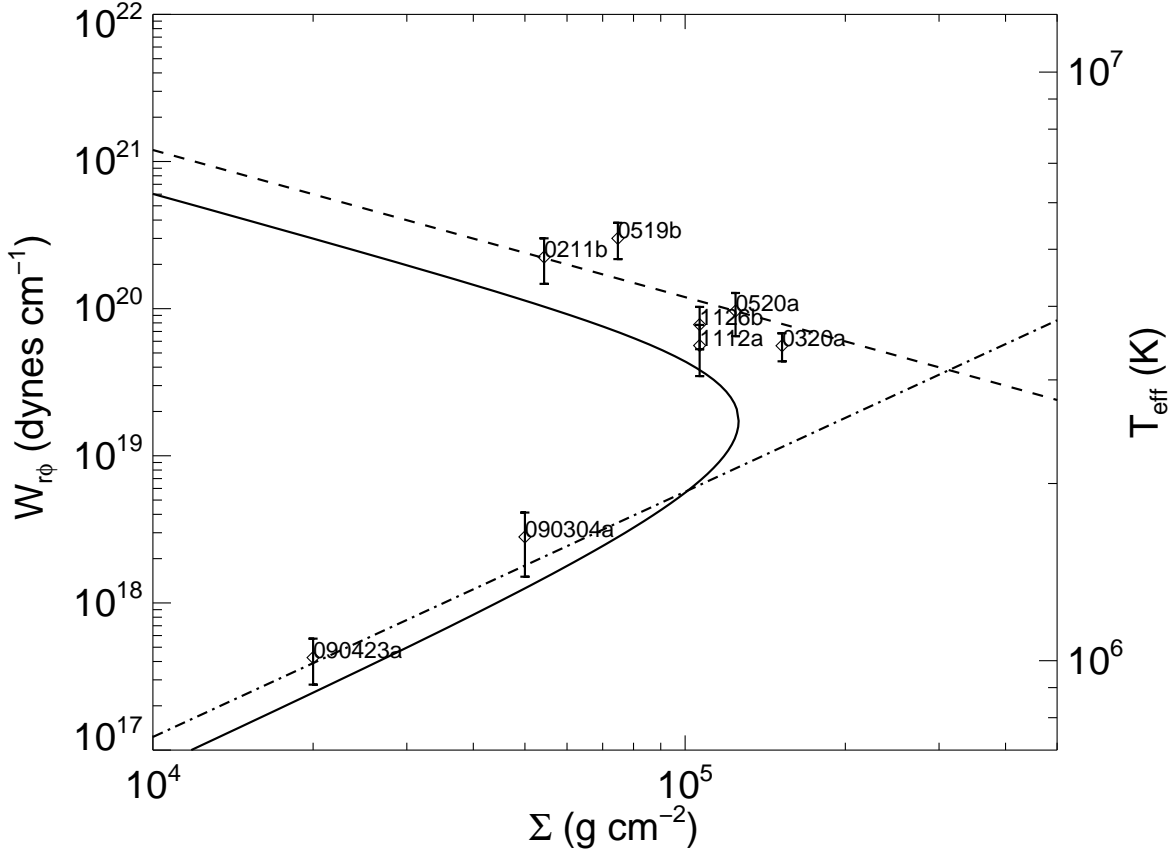}
\caption{Time averaged, vertically integrated stress as a function of surface
mass density for each simulation.  The right hand axis shows the corresponding
effective temperature of the radiation leaving each vertical
face of the box.  The solid curve shows the prediction of the vertically
integrated alpha disk model with internal structure parameters averaged over
all the simulations.  The dashed line shows the prediction of the radiation
pressure dominated alpha disk model with internal structure parameters averaged
over the radiation pressure dominated simulations.  The dot-dashed line shows
the prediction of the gas pressure dominated alpha disk model with internal
structure parameters measured from the gas pressure dominated simulations.
(See Appendix A for the equations used to define the internal
structure parameters.)
\label{fig:stresssigma}}
\end{figure}

For each simulation, we computed the vertically integrated stress as a function
of time, and then time averaged this over the simulation's duration, again
ignoring the first ten orbits.  The
results are plotted as a function of surface density in Figure
\ref{fig:stresssigma}.  Vertical error bars indicate the standard deviations
of the instantaneous fluctuations in stress about the mean.  (Due to mass
loss and an imposed density floor, the surface density also fluctuates, but
by at most two percent in all the simulations.)  The right hand axis indicates
the effective temperature of the radiation leaving each vertical face of
the box if perfect thermal equilibrium held.  We have also computed versions
of the diagram by time-averaging the radiation flux leaving both faces
of the box and computing the effective temperature, and time-averaging
the volume-integrated dissipation rate.  All three versions are almost
identical, as they must be, given the approximate thermal equilibrium that
has been established in each simulation.

The curves in Figure \ref{fig:stresssigma} show the results predicted by
standard alpha disk models with stress scaling with total pressure, but
with internal structure parameters based on the vertical structure observed
in the simulations themselves.  These parameters are defined in Appendix
A, and listed in Table \ref{vertstructuredata}.
The simulation data are clearly consistent with the alpha disk model, although
there are variations in the internal structure parameters that, when averaged
over all the simulations, produce an alpha disk curve that misses many of the
data points (solid curve in Figure \ref{fig:stresssigma}).  Note that the
alpha disk model predicts a maximum value of surface density $\Sigma_{\rm crit}$
above which an alpha disk cannot be in thermal equilibrium.  We have
not attempted to simulate such high surface densities.

The negative slope of the stress-surface density relation in the radiation
pressure dominated simulation data indicates a negative mass diffusion
coefficient, suggesting inflow instability if local thermal equilibrium is
everywhere maintained on the time scales associated with nonlocal, radial
mass transport.  Because this would be a diffusive instability, a perturbation
with radial wavelength $\lambda$ would have a characteristic growth rate
$\sim \alpha\Omega (H/\lambda)^2$ \citep{lig74b,sha76}, hence growing faster at
shorter wavelengths.  However, there must be a short wavelength
cutoff to this trend.  \citet{lig74b} argues that this cutoff is of order
the vertical scale height $H$ because of turbulent mixing on such length scales.
\citet{sha76} reach approximately the same conclusion based on the fact that
the linear thermal and inflow unstable modes of the alpha disk equations become
degenerate and then stable for radial wavelengths less than of order the
disk scale height.  This latter fact suggests that the short wavelength
cutoff to the inflow instability might actually be considerably larger than
the disk scale height, because the radiation pressure driven thermal
instability itself does not manifest itself in real MRI turbulence.
Insofar as thermal effects become increasingly important to the inflow
instability at short wavelengths, this may make the cutoff wavelength longer.

Because the radial extent of our simulation
domain is so small (significantly less than a scale height in all cases
- see Table \ref{simparams}), we do not expect
(and do not find) the inflow instability to be present in our
simulations.\footnote{Because the thermal instability of the alpha disk
also has a short radial wavelength cutoff, the reader might wonder how we
can claim with confidence that the thermal instability does not exist based
only on shearing box simulations with narrow radial domains \citep{hir09}.
This is because an infinite radial wavelength mode (one that does not
vary at all in radius) can still be present in such boxes, and would still
grow if it were truly thermally unstable.  Physically, the pure thermal
instability of the standard alpha disk arises because the vertically integrated
heating rate is more temperature sensitive than the cooling rate, and radial
variations are irrelevant.  In contrast, the growth rate of
infinite radial wavelength modes on the inflow instability branch is zero.}
In principle, vertically
stratified shearing boxes with much wider radial domains could produce
the instability, even though the shearing box has no net radial accretion.
In spite of our use of the term ``inflow'' instability, the instability is
really one of negative diffusion, and can grow even in the
absence of a net mass accretion flow through the box.  We can demonstrate
this fact explicitly by vertically integrating the shearing box equations (see
Appendix B).  This results in a radial mass diffusion equation,
\begin{equation}
{\partial\Sigma\over\partial t}={1\over(2-q)\Omega} {\partial^2\over
\partial x^2} W_{xy},
\end{equation}
identical to the local limit of equation (\ref{eqdsigmadt}), where
$q\equiv-d\ln\Omega/d\ln r$ and $|x|\ll r$.
If $\partial W_{xy}/\partial\Sigma<0$ under conditions
of thermal equilibrium, we would then expect inflow instability.  Stratified
shearing box simulations sufficiently wide in the radial direction
might therefore manifest this instability in the radiation pressure
dominated regime.  To see this effect may require radial box widths much larger than
the vertical pressure scale height, in order that different radial regions
be thermally decoupled and that stress scale with local pressure on
time scales longer than the local thermal time.  In any case, such simulations
would enable a determination of whether the inflow instability is real,
and what its minimum and fastest growing wavelengths are.

If the instability does manifest itself in MRI turbulence, such simulations 
would also help determine its nonlinear outcome.  Of course, a number of
hydrodynamic simulations have been done of the combined thermal/inflow
instability of radiation dominated alpha-disks (e.g. \citealt{hon91,szu98}).
However, the time evolution of these simulations is dominated by the
faster thermal instability, which we now know to be fictitious.
As far as we are aware, the only simulations that have ever been attempted of
the radiation dominated inflow instability on its own were
done by \citet{lig74b} himself, who numerically solved the alpha disk mass
diffusion equation, assuming that thermal equilibrium is strictly
maintained.
All his simulations therefore started with a surface density less than
$\Sigma_{\rm crit}$.  The resulting evolution rapidly produced clumping
of the surface density up to $\Sigma_{\rm crit}$ with optically thin rarefied
regions in between.  Both of these conditions violated the assumptions
on which the mass diffusion equation is based, and the numerical calculation
had to be stopped.  As \citet{lig74a} pointed out, surface densities
exceeding $\Sigma_{\rm crit}$ cannot be in thermal equilibrium within the
assumptions of the alpha model, because
heat generation always exceeds cooling through vertical radiative diffusion.
On the other hand, if the characteristic radial size of the clumps is as
small as the disk scale height, radial heat transport is likely to be
important in the nonlinear outcome.  Radiation MHD simulations with radially
wide shearing boxes could address all these issues.

Radiation MHD simulations could also clear up another question clouding
prediction of the outcome of this putative instability:
the negative slope of the stress-surface density relation on the radiation dominated
branch implies growing surface density fluctuations only if local
thermal equilibrium with purely vertical heat flow is maintained everywhere
while the instability tries to develop.  It could be that subtleties in the
behavior of MRI turbulence preclude this from happening, just as the
thermodynamics of the turbulence prevented the thermal instability from
manifesting itself.

This issue is closely related to the question
of exactly what the gas-dominated and radiation-dominated branches of equilibria
in Figure~\ref{fig:stresssigma} truly represent.  If this were a low-dimensional
dynamical system like the standard alpha-disk, then the fact that both branches
are thermally stable would imply an unstable equilibrium branch in between.
However, a stratified shearing box with real turbulence is not a
low-dimensional dynamical system---the spatial dependence of gas density,
pressure, velocity, magnetic field, and radiation pressure all influence its
evolution, and Figure~\ref{fig:stresssigma} is
probably better viewed as a projection of a very complicated dynamical phase
space.  In our experience, simulations that are initialized far away from
the equilibrium branches do not undergo steady heating or cooling, but instead
wander chaotically due to the fluctuating character of the turbulence.
Whether there is a true unstable thermal equilibrium branch between the
gas and radiation-dominated branches would be very well masked by this stochasticity
even if the vertically-integrated stress were the only significant dynamical variable;
the much higher dimensionality of the real system makes it essentially impossible to
test whether such a branch exists on the basis of simulation data.  It is conceivable
that a local perturbation in surface density would require considerable
time to reach the thermal equilibrium branch, and would in any case fluctuate
about that branch.  It is an open question whether or not this evolution
would be conducive to inflow instability.

It is also noteworthy that hydrostatic and radiative equilibrium necessarily
enforce a characteristic stress on the radiation dominated branch (equation
\ref{eqtaurphi}; \citealt{sha76}).  There is no such constraint on the gas
dominated branch, and in fact hydrostatic equilibrium is not even needed
to derive the relation between stress and surface density in the alpha model
on this branch (see Appendix A).  It is possible that these differences may
also be relevant to evolution on the inflow time scale in the gas and
radiation pressure dominated regimes in real MRI turbulence.

\section{Conclusions}

We have completed vertically stratified, local radiation MHD simuations of
magnetorotational turbulence with fixed vertical gravity over a range of
surface mass densities.  All of the simulations reach a thermal equilibrium,
but with continued long term fluctuations in the internal energy content.
We have confirmed earlier work \citep{tur04,hir09} that the radiation
pressure dominated thermal instability predicted by the standard alpha disk
model does not exist, even though the the box- and time-averaged radiation
to gas pressure ratios in the new simulations are as high as 70.

However, we also find that, when averaged over many thermal times, the vertically
integrated total thermal pressure (i.e., radiation plus gas pressure)
is well-correlated with the vertically integrated
stress.  Neither the vertically integrated gas pressure nor the geometric mean
of gas and radiation pressure exhibit such a good correlation.  The same simulation
data therefore yield a thermal equilibrium relation between surface density and
(long) time-averaged vertically integrated stress that follows the one predicted
by the traditional alpha model:
$\langle W_{r\phi}\rangle \propto \Sigma^{-1}$.  Consequently, if thermal
equilibrium is maintained on the long time scales associated with radial
mass transport by the turbulent stresses, these local results suggest
that the nonlocal clumping of mass associated with
the classic \citet{le74} inflow instability might develop.  Radiation
MHD shearing box simulations with much wider boxes in the radial direction,
or well-resolved global simulations, will be necessary to investigate this possibility.

This work was partially supported by a Grant-in-Aid for Scientific Research
(No. 20340040) from the Ministry of Education, Culture, Sports, Science and
Technology of Japan; and by NSF grants AST-0707624 and AST-0507455.
Numerical computations were carried out partially on the Cray XT4 at
the Center for Computational Astrophysics at the National
Astronomical Observatory of Japan, and on the SX8 at the Yukawa Institute
for Theoretical Physics at Kyoto University.
We thank Steve Balbus, Shane Davis, Chris Done, Jeremy Goodman,
John Hawley, Phil Marshall,
Gordon Ogilvie, John Papaloizou, Eli Rykoff, Chris Reynolds, and Jim Stone
for very useful conversations.  We also thank the anonymous referee for
helpful comments that improved the manuscript.

\appendix
\section{The Stress/Surface Density Relation for Alpha-Disks Accounting for
Internal Structure from the Simulations}

We derive one-zone vertical structure equations for an alpha-disk in terms
of parameters measured from the simulations as follows.
Vertical hydrostatic equilibrium implies that the midplane pressure is given
by
\begin{equation}
P(0)={1\over2}\Omega^2\Sigma H_{\rho1},
\label{eqhydrostatic}
\end{equation}
where $H_{\rho1}$ is a density scale height defined through the first vertical
moment of the density distribution,
\begin{equation}
H_{\rho 1}\equiv{2\over\Sigma}\int_{0}^\infty \rho(z) zdz.
\label{eqhrho1}
\end{equation}
The midplane pressure in the simulations is always dominated by
gas and radiation pressure,
\begin{equation}
P(0)={1\over3}aT(0)^4+{\Sigma kT(0)\over2\mu H_{\rho0}},
\label{eqp0}
\end{equation}
where $T(0)$ is the midplane temperature and
$\mu=0.6$ atomic mass units is the mean particle mass assumed in the
simulations.  We have eliminated the midplane density by defining
the zeroth vertical moment of the density distribution,
\begin{equation}
H_{\rho0}\equiv{\Sigma\over2\rho(0)}\equiv{1\over2\rho(0)}
\int_{-\infty}^\infty \rho(z)dz.
\label{eqhrho0}
\end{equation}

The fundamental assumption of an alpha-disk model is that the vertically
integrated stress $W_{r\phi}$ is $\alpha$ times the vertical integral of
the thermal pressure $P_{\rm th}$.  We may therefore write
\begin{equation}
W_{r\phi}=2\alpha H_PP(0),
\label{eqwrphi}
\end{equation}
where the thermal pressure scale height has been defined as
\begin{equation}
H_P\equiv{1\over2P(0)}\int_{-\infty}^\infty P_{\rm th}(z)dz
\label{eqhp}
\end{equation}
Thermal equilibrium implies that the flux $F=\sigma T_{\rm eff}^4$ emerging
from each face of the disk is given by
\begin{equation}
F=-{1\over2}W_{r\phi}r{d\Omega\over dr}={3\over4}\Omega W_{r\phi}
\label{eqfwrphi}
\end{equation}

Finally, the fact that most of the accretion power ultimately escapes vertically
through radiative diffusion in the simulations motivates us to write the
emergent flux as
\begin{equation}
F=\xi{2acT(0)^4\over3\kappa\Sigma},
\label{eqraddiff}
\end{equation}
where the opacity is dominated by electron scattering in our simulations,
$\kappa\simeq0.33$.  We have introduced a parameter $\xi$ which is usually
taken to be approximately unity in alpha-disk models.  As we discuss below,
however, our measured values of $\xi$ in the simulation are substantially
greater than unity.  This is presumably due to the fact that the dissipation
profile peaks off the midplane, and that a non-negligible fraction of the
accretion power is transported away from the midplane by mechanical motions.

Equations (\ref{eqhydrostatic}), (\ref{eqp0}), (\ref{eqwrphi}),
(\ref{eqfwrphi}), and (\ref{eqraddiff}) can be combined to give the
relationship between emergent flux or vertically integrated stress and
surface density.  In the gas and radiation pressure dominated limits, the
result is
\begin{equation}
F={3\over4}\Omega W_{r\phi}=\cases{\left[{3^5\kappa\over2^9ac\xi}\left(
{\alpha\Omega k\over\mu}\right)^4\left({H_P\over H_{\rho0}}\right)^4
\right]^{1/3}\Sigma^{5/3},&if $P_{\rm gas}(0)\gg P_{\rm rad}(0)$;\cr
{4c^2\Omega\over3\alpha\kappa^2}\xi^2\left({H_{\rho1}\over H_P}\right)
\Sigma^{-1},&if $P_{\rm rad}(0)\gg P_{\rm gas}(0)$.\cr}
\label{eqscurvelimits}
\end{equation}
This gives the usual result that the alpha disk is inflow stable if
gas pressure dominated, but inflow unstable if radiation pressure dominated.
Note that the hydrostatic equilibrium equation (\ref{eqhydrostatic}) is not
needed to derive the gas pressure dominated relation.

If neither pressure dominates at the midplane, the following equation can
be used to derive the relationship between flux or stress and surface
density:
\begin{equation}
5\tilde{F}^{3/4}-2\tilde{\Sigma}^{5/4}-3\tilde{F}^{5/4}\tilde{\Sigma}^{1/2}=0,
\label{eqscurve}
\end{equation}
where $\tilde{\Sigma}\equiv\Sigma/\Sigma_{\rm crit}$,
$\tilde{F}\equiv F/F_{\rm crit}$,
\begin{equation}
\Sigma_{\rm crit}=\left[{2^{19}\over(9)5^{10}}{a\over\Omega}
\left({c\xi\over\alpha\kappa}\right)^7\left({\mu\over k}\right)^4{H_{\rho1}^3
H_{\rho0}^4\over H_P^7}\right]^{1/8}
\label{eqsigmacrit}
\end{equation}
and
\begin{equation}
F_{\rm crit}={12c^2\xi^2\Omega\over25\alpha\kappa^2\Sigma_{\rm crit}}\left(
{H_{\rho1}\over H_P}\right).
\end{equation}
Apart from the extra dimensionless factors of $\xi$ and the pressure and
density scale heights, equation (\ref{eqsigmacrit}) agrees with the expression
for the maximum surface density consistent with thermal equilibrium that was
first derived by \citet{lig74a}.

Equations (\ref{eqscurvelimits}) and (\ref{eqscurve}) are what we use to plot
the alpha disk predictions in Figure~\ref{fig:stresssigma}.  We measured the
dimensionless parameters
describing the internal structure of the disk as follows.  Using the time
series of horizontally-averaged
data from each simulation, we measured $\alpha$ as the ratio of the vertically
integrated stress to vertically-averaged pressure, and $\xi$ in terms of the
ratio of emerging flux (averaged over both faces of the disk) to midplane
radiation energy density.  We then averaged both of these quantities over time.
We measured the scale height ratios $H_{\rho0}/H_{\rm P}$ and
$H_{\rho1}/H_{\rm P}$ from the time and horizontally-averaged vertical
profiles of density and thermal pressure from each simulation.  The results are
shown in Table \ref{vertstructuredata}.

The scale height ratio parameters are remarkably constant across all the
simulations.  The numerical values of these parameters are close to what
one would obtain from simple analytic equilibria in a gravitational field
that increases linearly with height.  An $n=3$ polytrope (adiabatic, radiation
pressure dominated) has $H_{\rho0}/H_P=1.125$ and $H_{\rho1}/H_P\simeq0.67$,
while an $n=3/2$ polytrope (adiabatic, gas pressure dominated) has
$H_{\rho0}/H_P=1.2$ and $H_{\rho1}/H_P\simeq0.69$.  Note that both ratios
are slightly larger in the gas pressure dominated polytrope, in agreement
with the trend that we see in the simulations.  These simple models
do not agree exactly with the simulations due in part to nonzero entropy
gradients.  In the simulations, the entropy generally increases away from
the midplane, so that the density must decrease faster
for a given pressure decrease compared to an adiabatic profile.  Hence
the ratio of density scale height to pressure scale height is smaller
than that of an adiabatic profile, and this is why the ratios we measure in the
simulations are slightly smaller than the polytropic ratios.

In contrast to the scale height ratios,  the stress parameter $\alpha$
shows a little more scatter.  The parameter $\xi$ clearly increases
as the simulations become more radiation pressure dominated.

\section{The Radial Mass Diffusion Equation for the Shearing Box}

The radial mass diffusion equation can be derived from the equations of the
shearing box \citep{haw95} as follows.  Define the azimuthal averaged surface
density at some radius $x$ and time $t$ as
\begin{equation}
\Sigma(x,t)\equiv{1\over L_y\Delta x}\int_{x-\Delta x/2}^{x+\Delta x/2}
dx^\prime\int_{-L_y/2}^{L_y/2}dy\int_{-L_z/2}^{L_z/2}dz \rho(x^\prime,y,z,t),
\end{equation}
and the mass-weighted vertical and azimuthal average of some quantity $X$ as
\begin{equation}
<X>(x,t)={1\over L_y\Delta x\Sigma}\int_{x-\Delta x/2}^{x+\Delta x/2}dx^\prime
\int_{-L_y/2}^{L_y/2}dy\int_{-L_z/2}^{L_z/2}dz\rho(x^\prime,y,z,t)
X(x^\prime,y,z,t).
\end{equation}
Here we have also performed a radial average over a length scale $\Delta x$
of order an assumed radial coherence length of the turbulence, which is
presumably of order the disk scale height.  Apart from the use of Cartesian
coordinates in a box, these definitions are exactly the same as the
vertical integrations and averages used to derive the alpha-disk
equations from the MHD equations by \citet{bal99}.  Applying this
averaging and vertical integration to the mass continuity and $y$-momentum
equations of the shearing box, we obtain
\begin{equation}
{\partial\Sigma\over\partial t}+{\partial\over\partial x}(\Sigma<v_x>)=0
\label{eqcontsb}
\end{equation}
and
\begin{equation}
{\partial\over\partial t}(\Sigma<v_y>)+{\partial\over\partial x}
(\Sigma<v_x><v_y>)+{\partial W_{xy}\over\partial x}=-2\Omega\Sigma<v_x>,
\label{eqymomsb}
\end{equation}
where $W_{xy}$ is the vertically-integrated and azimuthally-averaged Maxwell
and Reynolds stress,
\begin{equation}
W_{xy}(x,t)={1\over L_y}\int_{-L_y/2}^{L_y/2}dy\int_{-L_z/2}^{L_z/2}dz
\left(-{B_xB_y\over4\pi}+\rho v_x\delta v_y\right).
\end{equation}
Radial derivatives are defined through
differences over the length scale $\Delta x$.  We have also assumed zero mass
and Poynting flux through the vertical boundaries, and $<\delta v_y>=0$.
Equations (\ref{eqcontsb}) and (\ref{eqymomsb}) can then be combined into
a radial mass diffusion equation,
\begin{equation}
{\partial\Sigma\over\partial t}={1\over(2-q)\Omega} {\partial^2\over
\partial x^2} W_{xy},
\end{equation}
where $q\equiv-d\ln\Omega/d\ln r$ is the shear parameter.

%
%

\vfill\eject

%
%

\begin{deluxetable}{cccccc}
\tablecaption{Simulation Parameters \label{simparams}}
\tablehead{
\colhead{Simulation} & \colhead{$\Sigma$} & \colhead{$H$} & \colhead{Duration} &
\colhead{Box Dimensions} & \colhead{Grid Zones} \\
 & \colhead{(g cm$^{-2}$)} &  \colhead{(cm)} & \colhead{(orbits)} &
\colhead{($L_x/H\times L_y/H\times L_z/H$)} &
\colhead{($N_x\times N_y\times N_z$)}
}
\startdata
0211b & $5.43\times10^4$ & $5.83\times10^6$ & 264 &
$0.3375\times1.35\times6.3$ & $48\times96\times896$ \\
0519b & $7.48\times10^4$ & $4.37\times10^6$ & 403 &
$0.3375\times1.35\times6.3$ & $48\times96\times896$ \\
1112a & $1.06\times10^5$ & $1.46\times10^6$ & 610 &
$0.45\times1.8\times8.4$ & $48\times96\times896$ \\
1126b & $1.06\times10^5$ & $1.46\times10^6$ & 611 &
$0.45\times1.8\times8.4$ & $48\times96\times896$ \\
0520a & $1.24\times10^5$ & $1.17\times10^6$ & 603 &
$0.54\times2.16\times10.08$ & $48\times96\times896$ \\
0320a & $1.52\times10^5$ & $7.28\times10^5$ & 426 &
$0.6\times2.4\times11.2$ & $48\times96\times896$ \\
090304a & $5.00\times10^4$ & $3.13\times10^5$ & 600 &
$0.625\times2.5\times10$ & $32\times64\times512$ \\
090423a & $2.00\times10^4$ & $2.10\times10^5$ & 262 &
$0.5\times2.0\times8.0$ & $32\times64\times512$ \\
\enddata
\end{deluxetable}

\begin{deluxetable}{cccccc}
\tablecaption{Internal Structure Parameters Measured from Simulations
\label{vertstructuredata}}
\tablehead{
\colhead{Simulation} & \colhead{$\Sigma$} & \colhead{$\alpha$} &
\colhead{$\xi$} & \colhead{$H_{\rho0}/H_{\rm P}$} &
\colhead{$H_{\rho1}/H_{\rm P}$}\\
& \colhead{(g cm$^{-2}$)} & & & &
}
\startdata
0211b & $5.43\times10^4$ & $0.019\pm0.005$ & $5.4\pm1.2$ & 0.952 & 0.629\\
0519b & $7.48\times10^4$ & $0.015\pm0.004$ & $5.6\pm1.3$ & 0.949 & 0.630\\
1112a & $1.06\times10^5$ & $0.023\pm0.007$ & $4.5\pm1.1$ & 0.934 & 0.624\\
1126b & $1.06\times10^5$ & $0.020\pm0.006$ & $4.7\pm 1.0$ & 0.948 & 0.632\\
0520a & $1.24\times10^5$ & $0.016\pm0.006$ & $4.6\pm 1.0$ & 0.926 & 0.629\\
0320a & $1.52\times10^5$ & $0.015\pm0.004$ & $4.1\pm0.8$ & 0.990 & 0.645\\
090304a & $5.00\times10^4$ & $0.028\pm0.009$ & $2.8\pm0.6$ & 1.090 & 0.680\\
090423a & $2.00\times10^4$ & $0.028\pm0.007$ & $2.3\pm0.4$ & 1.127 & 0.685\\
\tableline
Mean$_{\rm rad}$\tablenotemark{a} &  & $0.017\pm0.002$ & $4.7\pm0.4$ & 0.950 & 0.631\\
Mean$_{\rm gas}$\tablenotemark{b} &  & $0.028\pm0.006$ & $2.5\pm0.4$ & 1.109 & 0.683\\
Mean$_{\rm all}$\tablenotemark{c} &  & $0.018\pm0.002$ & $3.4\pm0.3$ & 0.990 & 0.644\\
\enddata
\tablenotetext{a}{Averaged over the six radiation pressure dominated
simulations, i.e. excluding 090304a and 090423a.}
\tablenotetext{b}{Averaged over the the gas pressure dominated simulations
090304a and 090423a.}
\tablenotetext{c}{Averaged over all eight simulations.}
\end{deluxetable}


\begin{thebibliography}{}
\bibitem[Balbus \& Hawley(1998)]{bal98}Balbus, S. A., \& Hawley, J. F. 1998,
Rev. Mod. Phys., 70, 1
\bibitem[Balbus \& Papaloizou(1999)]{bal99}Balbus, S. A., \& Papaloizou,
J. C. B. 1999, ApJ, 521, 650
\bibitem[Belloni et al.(1997)]{bel97}Belloni, T., M\'endez, M., King, A. R.,
van der Klis, M., \& van Paradijs, J. 1997, ApJ, 479, L145
\bibitem[Blaes, Hirose, \& Krolik(2007)]{bla07}Blaes, O., Hirose, S., \& Krolik,
J. H. 2007, ApJ, 664, 1057
\bibitem[Burm(1985)]{bur85}Burm, H. 1985, A\&A, 143, 389
\bibitem[Done, Wardzi\'nski, \& Gierli\'nski(2004)]{don04}Done, C.,
Wardzi\'nski, G., \& Gierli\'nski, M. 2004, MNRAS, 349, 393
\bibitem[Hawley, Gammie, \& Balbus(1995)]{haw95} Hawley, J. F., Gammie, C. F.,
\& Balbus, S. A. 1995, ApJ, 440, 742
\bibitem[Hawley \& Krolik(2001)]{hk01} Hawley, J.F. \& Krolik, J.H. 2001, ApJ 548, 348
\bibitem[Hawley \& Krolik(2002)]{hk02} Hawley, J.F. \& Krolik, J.H. 2002, ApJ 566, 164
\bibitem[Hirose, Krolik, \& Stone(2006)]{hir06}Hirose, S., Krolik, J. H.,
\& Stone, J. M. 2006, ApJ, 640, 901
\bibitem[Hirose, Krolik, \& Blaes(2009)]{hir09}Hirose, S., Krolik, J. H., \&
Blaes, O. 2009, ApJ, 691, 16
\bibitem[Honma, Matsumoto, \& Kato(1991)]{hon91}Honma, F., Matsumoto, R.,
\& Kato, S. 1991, PASJ, 43, 147
\bibitem[Krolik, Hirose, \& Blaes(2007)]{kro07}Krolik, J. H., Hirose, S.,
\& Blaes, O.  2007, ApJ, 664, 1045
\bibitem[Lasota(2001)]{las01}Lasota, J.-P. 2001, New Astron. Rev., 45, 449
\bibitem[Lightman(1974a)]{lig74a}Lightman, A. P. 1974a, ApJ, 194, 419
\bibitem[Lightman(1974b)]{lig74b}Lightman, A. P. 1974b, ApJ, 194, 429
\bibitem[Lightman \& Eardley(1974)]{le74}Lightman, A. P., \& Eardley, D. M.
1974, ApJ, 187, L1
\bibitem[Lynden-Bell \& Pringle(1974)]{lyn74}Lynden-Bell, D., \& Pringle, J.
E. 1974, MNRAS, 168, 603
\bibitem[Merloni(2003)]{mer03}Merloni, A. 2003, MNRAS, 341, 1051
\bibitem[Merloni \& Fabian(2002)]{mer02}Merloni, A., \& Fabian, A. C. 2002,
MNRAS, 332, 165
\bibitem[Ohsuga et al.(2009)]{ohs09}Ohsuga, K., Mineshige, S., Mori, M.,
\& Kato, Y. 2009, PASJ, in press
\bibitem[Sakimoto \& Coroniti(1981)]{sak81}Sakimoto, P. J., \& Coroniti, F. V.
1981, ApJ, 247, 19
\bibitem[Sakimoto \& Coroniti(1989)]{sak89}Sakimoto, P. J., \& Coroniti, F. V.
1989, ApJ, 342, 49
\bibitem[Shakura \& Sunyaev(1973)]{sha73}Shakura, N. I., \& Sunyaev, R. A.
1973, A\&A, 24, 337
\bibitem[Shakura \& Sunyaev(1976)]{sha76}Shakura, N. I., \& Sunyaev, R. A.
1976, MNRAS, 175, 613
\bibitem[Shibazaki \& H\=oshi(1975)]{shi75}Shibazaki, N., \& H\=oshi, R. 1975,
Prog. Theor. Phys., 54, 706
\bibitem[Smak(1984)]{sma84}Smak, J. 1984, PASP, 96, 5
\bibitem[Stella \& Rosner(1984)]{ste84}Stella, L., \& Rosner, R. 1984,
ApJ, 277, 312
\bibitem[Szuszkiewicz(1990)]{szu90}Szuszkiewicz, E. 1990, MNRAS, 244, 377
\bibitem[Szuszkiewicz(1998)]{szu98}Szuszkiewicz, E., \& Miller, J. C. 1998,
MNRAS, 298, 888
\bibitem[Turner(2004)]{tur04}Turner, N. J., 2004, ApJ, 605, L45
\end{thebibliography}
\end{document}